# Privacy-preserving Pseudonym Schemes for Personalized 3D Avatars in Mobile Social Metaverses

1st Cheng Su, 2nd Xiaofeng Luo
School of Automation
Guangdong University of Technology
Guangzhou, China

3rd Zhenmou Liu
School of Systems Science and Engineering
Sun Yat-sen University,
Guangzhou 510275, China

4th *Jiawen Kang
School of Automation
Guangdong University of Technology
Guangzhou, China
*Corresponding author: kavinkang@gdut.edu.cn

5th Hao Min
Information Systems Technology and Design Pillar
Singapore University of Technology and Design
Singapore

6th Zehui Xiong
Information Systems Technology and Design Pillar
Singapore University of Technology and Design
Singapore

7th Zhaohui Yang, 8th Chongwen Huang
College of Information Science and Electronic Engineering
Zhejiang University
Hangzhou, China

*Abstract*—The emergence of mobile social metaverses, a novel paradigm bridging physical and virtual realms, has led to the widespread adoption of avatars as digital representations for Social Metaverse Users (SMUs) within virtual spaces. Equipped with immersive devices, SMUs leverage Edge Servers (ESs) to deploy their avatars and engage with other SMUs in virtual spaces. To enhance immersion, SMUs incline to opt for 3D avatars for social interactions. However, existing 3D avatars are typically generated through scanning the real faces of SMUs, which can raise concerns regarding information privacy and security, such as profile identity leakages. To tackle this, we introduce a new framework for personalized 3D avatar construction, leveraging a two-layer network model that provides SMUs with the option to customize their personal avatars for privacy preservation. Specifically, our approach introduces avatar pseudonyms to jointly safeguard the profile and digital identity privacy of the generated avatars. Then, we design a novel metric named Privacy of Personalized Avatars (PoPA), to evaluate effectiveness of the avatar pseudonyms. To optimize pseudonym resource, we model the pseudonym distribution process as a Stackelberg game and employ Deep Reinforcement Learning (DRL) to learn equilibrium strategies under incomplete information. Simulation results validate the efficacy and feasibility of our proposed schemes for mobile social metaverses.

## I.  INTRODUCTION

The continuous advancement in cutting-edge technologies like spatial computing, extended reality, and AI Generated Content (AIGC) has catalyzed the emergence of metaverses [1]. These advancements have accelerated the development of avatars, which serve as digital representations of humans within virtual spaces of metaverses. For example, the Apple Vision Pro headset [1] with their virtual avatar application named Persona, has garnered significant attention and success among the public, highlighting people's growing interest in personalized avatar creation within metaverses.

By merging mobile edge computing with social networks within metaverses, a novel paradigm known as the mobile social metaverse has emerged to cater to individuals' desires for engaging in immersive social activities (e.g., virtual tours and concerts) anytime and anywhere. Equipped with immersive devices like VR headsets, Social Metaverse Users (SMUs) can connect to edge servers (ESs) to access virtual spaces and interact with other SMUs through personally crafted avatars [2]. To enhance social experiences and presentation, SMUs rely on ESs to generate three-dimensional (3D) avatars, whose exquisite details further heighten the immersion and satisfaction of SMUs. For example, the Persona application on Apple Vision Pro can capture the real faces of SMUs to generate digital replicas as their virtual avatars in mobile social metaverses.

Despite the promising prospect of 3D avatars in mobile social metaverses, the associated privacy and security challenges should not be overlooked [3]. On one hand, since utilizing true faces or profile pictures for 3D avatar construction can pose privacy risks, SMUs tend to opt for animated characters, animals, or other images as their profile pictures to safeguard privacy on their social media like WeChat [4]. Hence, it becomes imperative to devise a highly decoupled method for 3D avatar generation, enabling SMUs to customize their avatars while ensuring profile anonymity in metaverses. On the other hand, given the social nature of mobile social metaverses, avatars may inadvertently expose the privacy of their real-world counterparts (i.e., SMUs) during extensive social interactions [3]. Therefore, it is also necessary to preserve the digital identity anonymity for SMUs in mobile social metaverses.

---
[1] https://www.apple.com/apple-vision-pro/

Fortunately, pseudonyms, serving as temporary credible identifiers, have demonstrated their efficacy in concealing the true identities of both physical entities (e.g., vehicles) in the Internet of Vehicles [3] and virtual entities (e.g., digital twins) in digital twin networks [5]. Therefore, employing pseudonyms to obscure the identities of avatars for SMUs offers a foundational solution for avatar privacy protection. Furthermore, leveraging edge computing technology, SMUs can request pseudonyms from Local Authorities (LAs) located in close proximity at the network edge to safeguard privacy [1], [3]. Nevertheless, the limited deployment of Edge Servers (ESs) in remote areas results in unprecedented network congestion due to the substantial and frequent requests for avatar updates and pseudonyms from SMUs. This congestion can considerably compromise privacy-preserving performance in mobile social metaverses. Hence, devising a feasible pseudonym resource allocation approach becomes imperative to ensure sustainable pseudonym distribution.

To this end, we present personalized 3D avatar construction framework in this paper. The key contributions are summarized as follows: 1) To cater to the diversity of SMUs' preferences for personalization and privacy protection in mobile social metaverses, we present a 3D avatar and avatar pseudonym generation framework, which incorporates a two-layer network model for personalized 3D avatar generation. With the generated avatars, the attribute-based pseudonyms are correspondingly generated and then distributed to SMUs for effective privacy protection. 2) We design a novel metric named Privacy of Personalized Avatar (PoPA) to evaluate the privacy levels of pseudonymous 3D avatars. Combined with the PoPA metric, we formulate a Stackelberg game between the LA and SMUs to foster sustainable pseudonym distribution. 3) Considering the variability of pseudonym demands and the information incompleteness of personalized 3D avatars generated by our devised network, we employ a Deep Reinforcement Learning (DRL) algorithm to solve the formulated game in mobile social metaverses.

## II. SYSTEM MODEL

### A. Overview of Mobile Social Metaverse

As shown in the upper of Figure. 1, we depict two use cases within mobile social metaverses, in which the SMUs who wear immersive devices living in the physical space can request avatars using personalized prompts from ESs to access virtual spaces, along with pseudonym demands for privacy protection. After deploying avatars, SMUs can socialize with others, such as participating in a virtual meeting, a virtual tour, and so on.

### B. Personalized 3D Avatar Construction

After the Stackelberg game reaches equilibrium, the LA located among trusted ESs can publish an optimal price for pseudonyms, while SMUs could correspondingly determine the number of pseudonyms to purchase (see the details in Sec. III-B). With the prompts, sensing data, and the final pseudonym demands, ESs can correspondingly execute edge inference tasks for personalized 3D avatar construction. Here, a two-layer 3D avatar construction framework is designed to achieve

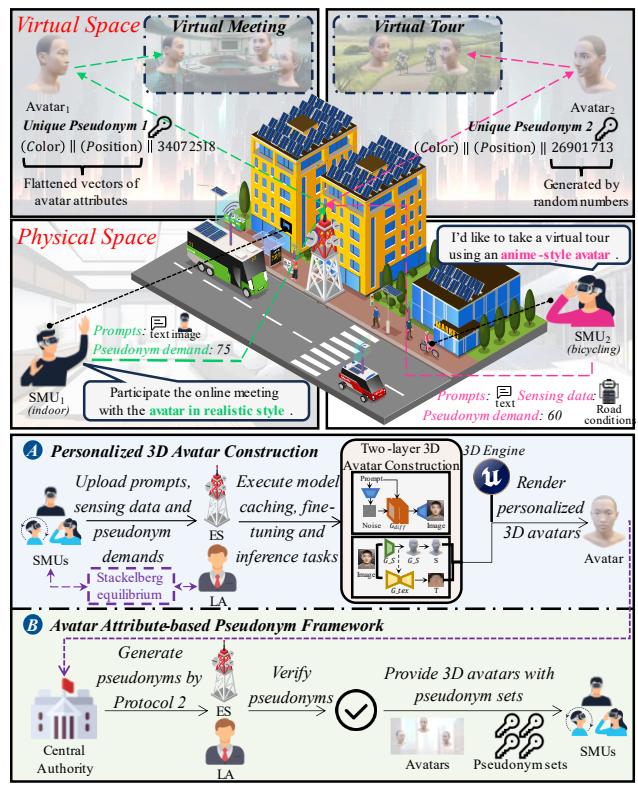

Figure 1. The workflow of privacy-preserving mobile social metaverses.

personalization, where the models used in each layer can be selected by SMUs.

The first layer in the framework is to generate profile images of avatars, where we use Stable Diffusion $G_{diff}$ [6] as the generative model. When the ES receives the text or image prompt $P_i$ along with the selected fine-tuned model $F_i$ to generate various styles of images (e.g., realism and animation styles) from $SMU_i$, $G_{diff}$ first initializes a latent image tensor and then the noise predictor can generate a predicted noise according to the input prompt and latent tensor. Then, the new latent image tensor is obtained by subtracting the predicted noise from the previous image. By iterating a specific number of iterations (i.e., sampling steps), the ES finally attains the profile image $X_i$, namely $X_i = G_{diff}(P_i, F_i)$.

The second layer is to generate 3D faces of avatars, where SMUs are allowed individually to choose different 3D shape regression model $G_S$ and texture map generation model $G_{tex}$ based on their personalized demands. With the output image $X_i$ from the first layer, the ES leverages $G_S$ with a 3D Morphable Model (3DMM) basis to regress the 3D shape in 3DMM coefficients and then construct 3D shape $S_i$. Meanwhile, $G_{tex}$ can generate a UV texture map $T_i$ based on 3DMM coefficients. These two processes are expressed as $S_i = G_S(X_i)$ and $T_i = G_{tex}(X_i, G_S(X_i))$, respectively. Combined with the generated 3D shape $S_i$ and UV texture map $T_i$, the ESs can construct a personalized 3D avatar $\mathbf{X}_i$ for $SMU_i$ through a 3D engine. The details can be seen in part. A of Figure. 1.

## C. Avatar Attribute-based Pseudonym Framework

To avoid the substantial communication overhead caused by frequent pseudonym requests from the SMUs, the LA often allocates a batch of pseudonyms to SMUs at once in the form of pseudonym sets. Let $r_i$ denote the number of pseudonyms requested by $SMU_i$. Upon receiving $SMU_i$'s pseudonym demand for the existing avatar or the regenerated avatar, the LA first extracts data $D_i$ that are related to the attributes of the generated 3D avatar, such as flattened feature vectors of lip color and position, and then forwards them to the Central Authority (CA). After auditing SMUs' identities, the CA initially generates $r_i$ random numbers. Based on the generated random numbers, CA calculates the pseudonymous part of pseudonyms by bilinear mapping [3]. By concatenating the attribute part and multiple pseudonym parts, the CA can generate an avatar pseudonym set containing $r_i$ pseudonyms for $SMU_i$. After a specific number of iterations, the CA returns all the pseudonym sets back to the LA. Thereafter, the LA will verify the validity of all avatar pseudonyms, and eventually provide renewed 3D avatars while distributing pseudonym sets to SMUs in mobile social metaverses. The attribute part of avatar pseudonyms can be used to conceal the real profile identities of virtual avatars, while the pseudonymous part can be used for digital identity anonymization, thus safeguarding privacy of associated SMUs. The SMUs can change their avatar pseudonyms along with other SMUs simultaneously, thereby eluding the persistent tracking by malicious adversaries in the mobile social metaverse. The detailed processes are presented in *part. B* of Figure. 1.

## III. PROBLEM FORMULATION

To investigate the privacy-preserving performance of our proposed schemes, we model the utilities of the LA and multiple SMUs and devise a Stackelberg game to facilitate pseudonym distribution processes. In this paper, we consider that a monopolistic LA and a set $\mathcal{I} = \{1, \ldots, i, \ldots, I\}$ of totally $I$ SMUs participate in the pseudonym distribution in mobile social metaverses.

### A. Utility Modeling

Similar to the definition of privacy entropy [3], given the 3D avatar shape $S_i$ and the UV texture map $T_i$, the PoPA of $SMU_i$ is calculated by:

$$H_i^T = -\log_2\left(\frac{S_i^{attr}}{S_i^{total}} + \frac{T_i^{attr}}{T_i^{total}} + \frac{1}{R_n^{R_l}}\right), \quad (1)$$

where $S_i^{total}$ and $T_i^{total}$ represent the total number of all data points of the generated 3D avatar shape and UV texture map of $SMU_i$, while $S_i^{attr}$ and $T_i^{attr}$ represent the number of attribute related data points which are used to generate pseudonyms. Since the number of attributes generated in different ways is almost equal, a larger total number means a lower probability of duplication of avatar pseudonyms, resulting in a higher PoPA and stronger privacy protection capacity. For the random pseudonymous part, $R_n$ and $R_l$ denote the range of optional numbers and the number of digits, respectively.

Referring to [1], when requesting $r_i$ pseudonyms and collectively changing with other avatars, $SMU_i$ can averagely increase the privacy level to $\overline{H_i} = \frac{\lambda_i}{\lambda_i+1}\left(1 + \frac{1}{\ln 2} - \frac{b_i \log_2 b_i - a_i \log_2 a_i}{b_i - a_i}\right) - 1$. Here, $\lambda_i$ is the pseudonym change frequency of $SMU_i$ in mobile social metaverses. The $a_i$ and $b_i$ denote the reciprocal of the maximum and minimum number of avatars at the social hotspot where $SMU_i$ resides, respectively [1]. Therefore, $SMU_i$ can benefit from requesting pseudonyms for privacy enhancement, whose utility function is expressed as

$$U_i^{PID} = \alpha_i \ln\left(1 + \left(H_i^T + \overline{H_i}r_i\right)\right) - pr_i, \quad (2)$$

where $\alpha_i$ denotes the utility coefficient of $SMU_i$'s privacy improvement by changing avatar pseudonyms, and $p$ denotes the unit selling price of avatar pseudonym by LA (the details are discussed below).

In addition to requesting pseudonyms from the LA, SMUs can choose whether to perform personalized avatar regeneration. Here, we employ a logarithmic function $\ln(\cdot)$ to quantify the satisfaction of SMUs for their personalized 3D virtual avatar renewals. Thus, the profit of $SMU_i$ for meeting his personalized 3D avatar demand can be calculated by

$$U_i^{Avatar} = x_i\left(\gamma_i \ln\left(\frac{\mu_i}{\mu_i^{th}} + \frac{\tau_i^{th}}{\tau_i}\right) - C_a\right). \quad (3)$$

Here, $x_i \in \{0,1\}$ is the binary variable denoting whether $SMU_i$ chooses to regenerate his unique avatar when requesting pseudonyms from the ES. The $\gamma_i$ is unit profit of $SMU_i$ for satisfying immersion by individually selecting two kinds of AI models. $\mu_i$ and $\mu_i^{th}$ are the CLIP Score [7] metric and the corresponding minimum threshold, of which the CLIP Score is a common metric used to reflect the correlation between text and image. $\tau_i$ and $\tau_i^{th}$ are the Learned Perceptual Image Patch Similarity (LPIPS) [8] metric and the corresponding maximum threshold. The LPIPS denotes the Euclidean distance of facial attributes between the input image and 2D rendered image from the generated 3D output, extracted by a deep learning model. $C_a$ is the fixed cost of regenerating a new avatar. $SMU_i$ both benefits by requesting pseudonyms for privacy enhancement and regenerating 3D avatars for personalization. Therefore, the total utility of $SMU_i$ is given as

$$U_i(r_i) = \omega_1 U_i^{PID} + \omega_2 U_i^{Avatar}, \quad (4)$$

where the weight factors $\omega_1$ and $\omega_2$ are used to characterize the proportion of privacy preservation and satisfaction for personalization, with $\omega_1 + \omega_2 = 1$.

On account of the information sensitivity of avatar profiles and pseudonyms, the trusted LA is considered the sole pseudonym distributor in mobile social metaverses to ensure pseudonym authenticity. To prevent SMUs from excessively requesting pseudonyms resulting in network congestion, the LA fixes a price $p$ for pseudonyms to limit the pseudonym purchase by SMUs. When selling $r_i$ pseudonyms to $SMU_i$, the utility associated with pseudonyms of LA is calculated by

$$U_{LA,i}^{PID} = pr_i - cr_i, \quad (5)$$

where $c$ is the unit cost of pseudonym transmission.

With the abundant computational and storage resources, the LA containing multiple ESs can accomplish large-scale edge inference tasks to accelerate 3D avatar generation tasks requested by SMUs [9]. To achieve the customization desires, SMUs are privileged to liberally select AI models for text-to-image and image-to-3D generation. If $SMU_i$ selects the models that are not cached beforehand, the LA should bear the model switch cost, which is calculated by $c_{i,m,n}^{switch} = \lambda_m \mathbf{1}(m_i \notin \mathcal{M}_i^{past}) + \lambda_n \mathbf{1}(n \notin \mathcal{N}_i^{past})$, where $m_i \in M$ and $n_i \in N$ are the selected model $m$ and model $n$ by $SMU_i$, with $M$ denoting the set of all optional styles of fine-tuned diffusion models for text-to-image generation (e.g., animation or realistic styles), while $N$ denoting the set of all optional models for 3D face reconstruction (e.g., StyleGAN2 [10]). $M^{past}$ and $N^{past}$ are the set of the selected model $m$ and model $n$ on the ES nearest to $SMU_i$, respectively. $\lambda_m$ and $\lambda_n$ represent the cost coefficient for caching the model $m$ and model $n$ [9], respectively. $\mathbf{1}(\cdot)$ is the indicator function. Besides, the transmission cost of the LA for input prompts and output inference results is represented by $c_{i,m,n}^{trans} = \kappa_i$ [9]. The computation cost can be calculated by $c_{i,m,n}^{comp} = \frac{g_m + g_n}{f}$, where $g_m$ and $g_n$ are the computational overhead for executing inference tasks of the model $m$ and model $n$, respectively. $f$ is the GPU computing capacity of ESs [9]. Therefore, when $SMU_i$ requests an avatar regeneration, the total avatar construction cost of the LA is represented as $c_{i,m,n}^{Avatar} = C_{i,m,n}^{switch} + C_{i,m,n}^{trans} + C_{i,m,n}^{comp}$. As $SMU_i$ requests a renewal of his avatar in the metaverse, the part associated with the avatar attribute of the pseudonym requires modifying accordingly. Hence, the utility of LA for providing avatar regeneration services to $SMU_i$ is calculated by

$$U_{LA,i}^{Avatar} = x_i \left( C_a - c_{i,m,n}^{Avatar} - C_l \Phi(D_i) \right), \quad (6)$$

where $C_t$ is the unit data transmission cost from the LA to the CA, and $\Phi(D_i)$ is the data size of $D_i$. Consequently, the overall utility function of LA is defined as

$$U_{LA}(p) = \sum_{i \in \mathcal{I}} \left( \eta_1 U_{LA,i}^{PID} + \eta_2 U_{LA,i}^{Avatar} \right), \quad (7)$$

where $\eta_1$ and $\eta_2$ are weight factors with $\eta_1 + \eta_2 = 1$.

*B. Stackelberg Game Formulation and Analysis*

Based on the above analyses, a monopolistic market has emerged, where the LA functioning as a monopolist has the right to determine the price of pseudonyms and SMUs can dynamically adjust their demands for pseudonyms in response to the ever-changing price. Given the utility functions of both the LA and SMUs, we formulate a Stackelberg game with LA as the leader and SMUs as the followers. For SMUs, the problem is formulated as

$$Problem 1: \quad \max_{r_i} U_i(r_i), \quad (8)$$

$$s.t. \quad r_i > 0.$$

For the LA, the problem is formulated as

$$Problem 2: \quad \max_p U_{LA}(p), \quad (9)$$

$$s.t. \quad 0 < \sum_{i \in \mathcal{I}} r_i \leq R^{max},$$

$$r_i > 0, \forall i \in \mathcal{I},$$

$$0 < c \leq p \leq P^{max}.$$

Here, $R^{max}$ and $P^{max}$ are the maximum number of pseudonyms that LA can provide and the maximum price per pseudonym.

**Theorem 1**. *The Stackelberg equilibrium of the formulated game between the LA and MUs is unique.*

*Proof.* The first-order and second-order derivatives of $U_i(r_i)$ are calculated by

$$\frac{\partial U_i(r_i)}{\partial r_i} = \frac{\alpha_i \overline{H_i}}{1 + H_i^T + \overline{H_i} r_i} - p, \quad (10)$$

$$\frac{\partial^2 U_i(r_i)}{\partial r_i^2} = -\frac{\alpha_i \overline{H_i}^2}{\left(1 + H_i^T + \overline{H_i} r_i\right)^2} < 0. \quad (11)$$

We can observe that the second-order derivative is negative while the first-order derivative has a unique zero point, indicating that the utility function of $U_i$ is strictly concave. With the first-order optimality condition (i.e., $\frac{\partial U_i(r_i)}{\partial r_i} = 0$) [2], the optimal number of pseudonyms requested by $SMU_i$ is expressed as $r_i^* = \left( \frac{\alpha_i}{p} - \frac{1}{\overline{H_i}} \right)(1 + H_i^T)$.

The above analyses signify that no matter how much LA sets the price $p$, SMUs can always adopt the optimal pseudonym purchasing strategy to maximize their respective utility. By substituting $r_i^*$ into Eq. (11), the first-order and second-order derivatives of $U_{LA}(p)$ are calculated by

$$\frac{\partial U_{LA}(p)}{\partial p} = \sum_{i \in \mathcal{I}} \left( (1 + H_i^T) \left( \frac{\alpha_i c}{p^2} - \frac{1}{\overline{H_i}} \right) \right), \quad (12)$$

$$\frac{\partial^2 U_{LA}(p)}{\partial p^2} = -\sum_{i \in \mathcal{I}} \left( (1 + H_i^T) \frac{2\alpha_i c}{p^3} \right) < 0. \quad (13)$$

We can also conclude that the utility function of LA is strictly concave similarly. By setting $\frac{\partial U_{LA}(p)}{\partial p} = 0$, the optimal pseudonym pricing strategy, namely, $p^* = \sqrt{\frac{c \sum_{i \in I} \alpha_i (1 + H_i^T)}{\sum_{i \in I} \left( \frac{1 + H_i^T}{\overline{H_i}} \right)}}$ can be adopted by the LA to maximize its utility.

Therefore, we discover that the leader LA can first achieve a unique Nash equilibrium by setting the optimal price $p^*$, and all the SMUs can also achieve a unique Nash equilibrium by taking the optimal pseudonym purchasing strategy with $r^* = \{r_i^*\}_{i=1}^I$. Consequently, the Stackelberg equilibrium of the formulated game between the LA and SMUs always exists and is unique [2].

## IV. A DRL-BASED SOLUTION FOR INCOMPLETE GAME

Due to privacy concerns, SMUs may not provide their sensitive information to the LA. In this way, although the Stackelberg game equilibrium exists, the LA still has difficulty in obtaining the optimal pricing strategy because of the information incompleteness. To address this, we transform the formulated game with incomplete information into a partially observable Markov decision process and then adopt a DRL method to solve it [1].

In the context of distributing pseudonyms to SMUs, the LA only possesses incomplete information about SMUs in practical applications. Specifically, the LA can solely observe the determined actions of SMUs over past $L$ rounds. During the current game round $k$, the partially observable state space encompasses two parts: the pricing strategy of LA and the pseudonym demands of SMUs over the past $L$ rounds, which is denoted as $o_k \triangleq \{p_{k-L}, r_{k-L}, p_{k-L+1}, r_{k-L+1}, ..., p_{k-1}, r_{k-1}\}$.

Acting as an intelligent agent for network optimization, the LA can take a pricing action $p_k$ based on the observation $o_k$ to maximize its utility. Hence, the LA's policy is represented as $\pi(p_k|o_k, \theta) \to [c, p_{max}]$, where $\pi_\theta$ represents a neural network utilized to learn the optimal pricing decision.

According to the current observation $o_k$, the learning-based LA adopts an action $p_k$ to gain reward, and then $o_k$ transitions to $o_{k+1}$ [1]. Therefore, the reward function is defined as

$$R(o_k, p_k) = \begin{cases} 1, U_{LA}^k \geq U_{max}^k, \\ 0, U_{LA}^k < U_{max}^k, \end{cases} \quad (14)$$

where $U_{LA}^k$ is the utility function of LA in the current round $k$, calculated by (11). $U_{max}^k$ is the maximum utility that LA has ever achieved in history [2].

## V. PERFORMANCE EVALUATION

In this section, we evaluate the quality of personalized 3D avatar generation and the performance of DRL-based pseudonym distribution. We describe the experimental settings, experimental results and analyses of the 3D avatar generation and pseudonym distribution in Sec. V-A and Sec. V-B.

### A. Quality of Personalized 3D Avatar Generation

We use Stable Diffusion [6] for image generation, in which two optional fine-tuned models can generate images in different styles according to the text and image prompts. For the parameter setting, the generative step, classifier-free guidance, and generative image size are set to 50, 7.5, and 512 × 512, respectively. For the 3D face generation, we use a Deep3D model [11] with the recent 3DMM basis HiFi3D++ [12] to regress the 3D shapes in 3DMM coefficients, while the UV texture maps are generated by pre-trained StyleGAN2 [10] model. The model inferences are self-supervised learning processes, optimizing all parameters by minimizing the LPIPS metric [8] mentioned in Sec. III.

Meanwhile, we compare our methods with another 3D face generation method *DreamFace* [13], to study the personalization and the security of generated avatar pseudonyms. The data related to eyes, ears, and noses are employed as avatar attributes for $SMU_1$. In terms of eye attributes, $S_i^{attr}$ and $T_i^{attr}$ are set to 100000 and 400, respectively. For the ear attributes, $S_i^{attr}$ and $T_i^{attr}$ are set to 95000 and 500, respectively. For the nose attributes, $S_i^{attr}$ and $T_i^{attr}$ are set to 150000 and 300, respectively. $R_n$ and $R_l$ are set to 9 and 4, respectively. $S_i^{total}$ and $T_i^{total}$ are obtained by different algorithms due to the quality distinction of different 3D face generation methods.

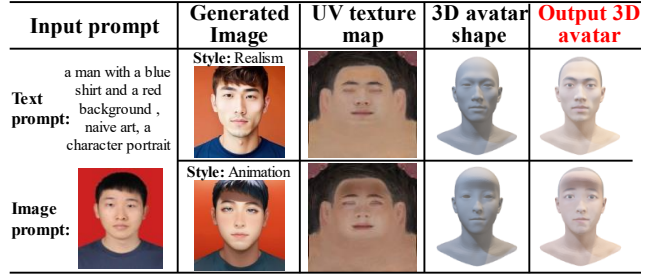

Figure 2. Example of our generated personalized 3D avatars.

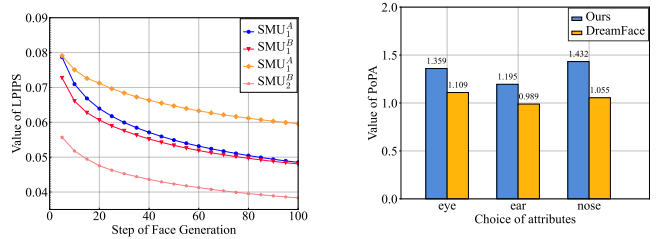

(a) LPIPS value versus steps of 3D face generation.

(b) PoPA value under different choices of avatar attributes.

Figure 3. Performance of the proposed attribute-based avatar pseudonym.

As shown in Figure. 2, after the SMU freely provides the text and image prompt, the nearest ES generates two kinds of avatar profile images based on model A (Realism style) and model B (Animation style), which are ultimately used for 3D avatar construction. We can find that, our proposed methods reserve high fidelity of the output 3D avatars while preventing 3D avatars from appearing in mobile social metaverses with the real faces of SMUs, thereby effectively protecting the profile identity privacy. Figure. 3(a) shows that no matter which model the two SMUs choose, the LPIPS value always decreases as the number of steps increases, meaning the strong suitability of our methods in different scenarios. Then, we use the proposed PoPA metric to compare the performance of our method with *DreamFace* on privacy security. In Figure. 3(b), we can find that our proposed algorithm achieves a higher PoPA compared to the *DreamFace* method. The above results demonstrate that our schemes can generate high-quality personalized 3D avatars while realizing better privacy-preserving performances.

### B. Performance Analysis of Pseudonym Distribution

For the DRL-based avatar pseudonym distribution optimization, we employ an on-policy Proximal Policy Optimization (PPO) algorithm with actor-critic framework as our method in line with the settings in [2]. We consider that 6 SMUs are evenly spread across 3 edge servers within LA coverage, and both the optional number of image generation model and 3D face generation model are 3. For the utility of $SMU_i$, the parameters $a_i, b_i, \alpha_i, \mu_i^{th}$, and $\tau_i^{th}$ are set to 1/160,

1/10, 15, 15, and 0.08, respectively. The parameters $\gamma_i$, $\mu_i$, and $\tau_i$ follow a uniform distribution, namely, $\gamma_i \sim U(1.5,2)$, $\mu_i \sim U(20,40)$, and $\tau_i \sim U(0.02,0.06)$. Referring to [3], the default average pseudonym change frequency of SMUs $\bar{\lambda}$ is set to 1.5. According to the numerical result in Figure. 3(b), the default

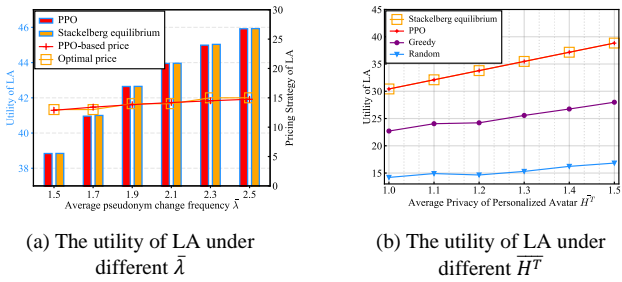

(a) The utility of LA under different $\bar{\lambda}$

(b) The utility of LA under different $\overline{H^T}$

Figure 4. Performance analyses of DRL-based avatar pseudonym distribution.

average PoPA is set to 1.5. For the utility of LA, the parameter $c$, $Ct$, $\Phi(D_i)$, $\kappa_i$, $\lambda_m$, $\lambda_n$, $f$, $g_m$, $g_n$ are set to 5, 0.2, 1, 0.05, 0.3, 3, 312000, 60000, 600000, respectively [9].

Figure. 4(a) exhibits the utility and pricing strategy of LA under different average pseudonym change frequency $\bar{\lambda}$. Focusing on the bar graph with blue edge lines, we can observe that as $\bar{\lambda}$ increases, the utility of LA also increases accordingly. The reason is that when SMUs change their avatar pseudonyms more frequently, their pseudonym demands also increase. Therefore, the LA can raise the price for pseudonyms to make more profits, which also can be seen in the solid line about the pricing strategy of LA in 4(a). Moreover, we can also find that both in terms of LA utility and pricing strategy, our DRL-based solution under incomplete information is close to Stackelberg equilibrium under complete information, demonstrating its feasibility in mobile social metaverses.

In Figure. 4(b), we provide the simulation results of the utility of LA under different average PoPA of SMUs $\overline{H^T}$. It is evident that the utility of LA increases as $\overline{H^T}$ under our DRL-based solution, which is almost equal to the Stackelberg equilibrium. Meanwhile, we can find that our DRL-based schemes outperform the baseline random and greedy algorithms by 27.2% and 55.7%, respectively.

## VI. CONCLUSION

In this paper, we presented a 3D avatar and pseudonym generation for personalized and privacy-preserving avatar socializing in mobile social metaverses. To achieve personalized avatar generation, we designed a double-layer 3D avatar construction framework. Based on this, we introduced the avatar pseudonym containing the avatar attributes to jointly protect profile and digital identity privacy for SMUs. We proposed a metric called PoPA to quantify the privacy-preserving degree of avatar pseudonyms. Subsequently, we studied the optimization problem of avatar pseudonym distribution between the LA and SMUs, formulating the resource trading issue as a Stackelberg game. Deep reinforcement learning (DRL) is applied to tackle this game under the condition of incomplete information. Finally, simulation results validate the security and efficacy of the proposed schemes. In the future, we will enhance the assessment of security and satisfaction for SMUs by incorporating more effective metrics related to privacy and social attributes.